
\magnification=1200
 \hsize=16.8 true cm
 \vsize=21.8 true cm
 \hoffset=-0.35 true in
 \baselineskip= 15pt plus 4pt

\def\a{\hat a}
\def\ap{\hat a^\prime}
\def\b{\hat b}
\def\bp{\hat b^\prime}
\def\AP{A^\prime}
\def\BP{B^\prime}

 \parindent=1.0cm
 \parskip=0.5cm
\pageno=1

 \vskip 1.5cm
 \centerline{NONLOCALITY AS AN AXIOM FOR QUANTUM THEORY\footnote{*}
{Talk presented at {\it 60 Years of E.P.R.}
(Workshop on the
Foundations of Quantum Mechanics, in honor of
Nathan Rosen),
Technion, Israel, March 1995}}
 \vskip 1.5cm
 \baselineskip= 10pt plus 4pt

 \centerline{Daniel Rohrlich and Sandu Popescu}
\medskip
\centerline{\it School of Physics and Astronomy, Tel-Aviv University}
\centerline{\it Ramat-Aviv, Tel-Aviv 69978 Israel}
 \vskip 1.25cm
 \centerline{ABSTRACT}
 \parindent=1.3cm
 \baselineskip 10pt plus 2pt
Quantum mechanics and relativistic causality together imply nonlocality:
nonlocal correlations (that violate the CHSH inequality) and nonlocal
equations of motion (the Aharonov-Bohm effect).  Can we invert the logical
order?  We consider a conjecture that nonlocality and relativistic
causality together imply quantum mechanics.  We show that correlations
preserving relativistic causality can violate the CHSH inequality more
strongly than quantum correlations.  Also, we describe nonlocal equations
of motion, preserving relativistic causality, that do not arise in quantum
mechanics. In these nonlocal equations of motion, an experimenter ``jams"
nonlocal correlations between quantum systems.
\vskip 0.25cm

\parindent=1.0cm
\baselineskip 15pt plus 2pt
\noindent 1. \hskip 0.5cm {INTRODUCTION}

Two aspects of quantum nonlocality are nonlocal correlations and nonlocal
equations of motion.  Nonlocal {\it correlations} arise in settings
such as the one discussed by Einstein, Podolsky and Rosen$^1$. As Bell$^2$
showed (and Aspect has reviewed in his lecture here) no theory of
local variables can reproduce these correlations. The Aharonov-Bohm
effect$^3$ is also nonlocal in that an electromagnetic field
influences an electron in a region where the field vanishes. The
field induces a relative phase between two sets of paths available
to an electron, displacing the interference pattern
between the two sets of paths.
Thus, the Aharonov-Bohm effect implies nonlocal
{\it equations of motion}.$^4$  Both aspects of quantum nonlocality
arise within nonrelativistic quantum theory.  However, the very
definition of a local
variable is relativistic:  a local variable can be influenced only by
events in its backward light cone, and can
influence events only in its forward light cone.  In this sense, quantum
mechanics and relativity {\it together} imply nonlocality.  They coexist
because quantum correlations preserve relativistic causality (i.e.
they do not allow us to transmit signals faster than light).
But quantum mechanics does not
allow us to consider isolated systems as separate, as Einstein, Podolsky
and Rosen$^1$ assumed. This violation of not the
letter but the spirit of special relativity has left many physicists
(including Bell) deeply unsettled. Today, quantum nonlocality seems as
fundamental---and as unsettling---as ever.  If nonlocality is
fundamental, why not make nonlocality an axiom of quantum
theory rather than a consequence?  Can we then
invert the logical order, showing that nonlocality and relativistic causality
together imply quantum theory?

\noindent 2. \hskip .5 cm NONLOCALITY I:  NONLOCAL CORRELATIONS
\nobreak

Quantum mechanics and relativistic
causality together give rise to nonlocal correlations, which many physicists
regard as a negative aspect
of quantum theory.  Here, we regard quantum nonlocality as a
positive aspect of quantum theory.  What new possibilities does
quantum nonlocality offer us? In particular, if we make nonlocality an
axiom, what becomes of the logical structure of quantum theory?$^{5-7}$  The
special theory of relativity can be deduced in its entirety
from two axioms:  the equivalence of inertial reference frames, and the
constancy of the speed of light. Aharonov$^7$ has proposed such
a logical structure for quantum theory.  Let us take, as axioms of quantum
theory, relativistic causality and nonlocality. As an initial, immediate
result, we deduce that quantum theory is not deterministic,
otherwise these two axioms would be incompatible.$^7$  Two ``negative"
aspects of quantum mechanics---indeterminacy and limits on
measurements---then appear as a consequence of a fundamental ``positive"
aspect:  the possibility of nonlocal action.  Moreover, by taking
nonlocality as an axiom, we free ourselves of the need to explain it.

     We have not yet defined the axiom of nonlocality.  Relativistic
causality is well defined, but quantum nonlocality arises
both in nonlocal correlations and in the Aharonov-Bohm effect. In this
section we consider nonlocal correlations.
We ask which theories yield nonlocal
correlations while preserving causality.
Our result is independent of quantum mechanics or any
particular model.  We find$^{8}$ that quantum mechanics is only one of a
class of theories consistent with our two axioms, and, in a certain sense,
not even the most nonlocal theory.

The Clauser, Horne, Shimony, and Holt$^{9}$ (CHSH) form of
Bell's inequality,
holds in any classical theory (that is, any theory of local hidden
variables).  It  states that a certain combination of correlations lies
between -2 and 2:
$$
-2 \le E(A,B)+E(A,\BP )+E(\AP ,B)-E(\AP ,\BP )\le 2
{}~~~~.\eqno(1)$$
Besides 2, two other numbers, $2\sqrt{2}$ and $4$, are
important bounds on the CHSH sum of correlations.  If the four correlations in
Eq. (1)
were independent, the absolute value of the sum could be as much as 4. For
quantum correlations, however,
the CHSH sum of correlations is bounded$^{10}$ in absolute value by
$2\sqrt{2}$. Where does this bound come from?  Rather than asking
why quantum correlations violate the CHSH inequality, we might ask why they do
not violate it {\it more}.

Let us say that of the two axioms proposed above, the axiom
of nonlocality implies that quantum correlations violate the CHSH
inequality at least sometimes.  We may then guess that the other axiom,
relativistic causality, might imply that quantum correlations do not violate it
maximally. Could it be that relativistic causality restricts the violation to
$2\sqrt {2}$ instead of 4?  If so, then the two axioms
determine the quantum violation of the CHSH inequality.
To answer this question, we ask what restrictions relativistic causality
imposes on joint probabilities. Relativistic causality forbids sending messages
faster than light. Thus, if one observer measures the observable $A$, the
probabilities for the outcomes $A=1$ and $A=-1$ must be independent of whether
the other observer chooses to measure $B$ or $\BP$. However, it can be
shown$^{8,11}$ that this constraint does not
limit the CHSH sum of quantum correlations to $2\sqrt{2}$.
For example, imagine a ``superquantum" correlation function $E$
for spin measurements
along given axes.  Assume $E$ depends only on the relative angle $\theta$
between axes. For any pair of axes, the outcomes $\vert \uparrow \uparrow
\rangle$ and $\vert \downarrow \downarrow \rangle$ are equally likely, and
similarly for $\vert \uparrow
\downarrow \rangle$ and $\vert \downarrow \uparrow \rangle$.  These four
probabilities sum to 1, so the probabilities for $\vert \uparrow \downarrow
\rangle$ and $\vert \downarrow \downarrow \rangle$ sum to $1/2$. In any
direction, the probability of $\vert \uparrow \rangle$ or $\vert \downarrow
\rangle$ is $1/2$ irrespective of a measurement on the other particle.
Measurements on one particle yield no information about measurements on the
other, so relativistic causality holds.
The correlation function then satisfies $E(\pi - \theta ) = -E(\theta )$.
Now let $E(\theta )$ have the form

     (i) $E (\theta ) =1$ for $0 \le \theta \le \pi /4$;

     (ii) $E(\theta )$ decreases monotonically and smoothly from 1 to -1
as $\theta$ increases from $\pi /4$ to $3\pi / 4$;

     (iii) $E(\theta) = -1$ for $3\pi /4 \le  \theta \le \pi$.

Consider four measurements along axes defined by unit vectors $\ap$, $\b$,
$\a$, and $\bp$ separated by successive angles of $\pi /4$ and lying in a
plane. If we now apply the CHSH inequality Eq. (1)
to these directions, we find that the sum of correlations
$$
E(\a , \b ) +E(\ap , \b )+E(\a , \bp )-E(\ap , \bp )
=3E(\pi /4 ) - E(3\pi /4) = 4
{}~~~~\eqno(2)$$
violates the CHSH inequality with the maximal value 4.  Thus, a
correlation function could satisfy relativistic causality and still violate
the CHSH inequality with the maximal value 4.
\goodbreak
\noindent 3. \hskip 0.5cm NONLOCALITY II:  NONLOCAL EQUATIONS OF MOTION
\nobreak

In one version of the Aharonov-Bohm effect, an isolated magnetic
flux, inserted between two slits, shifts the interference pattern of electrons
passing through the slits.  It thereby
affects the electron's momentum, since the
electron arrives at a different point than it would without the
electromagnetic field.  Thus, the Aharonov-Bohm effect implies nonlocal
{\it equations of motion}.$^4$ Aharonov has shown$^7$ that a physical quantity,
the {\it modular momentum} of the flux,$^{12}$ is uncertain exactly as
required to keep us from seeing a violation of causality.
In general,
modular momentum is measurable and obeys a nonlocal equation of motion.
But when the flux is located between the slits, its modular momentum
is completely uncertain.

     Is quantum mechanics the {\it only} relativistically causal theory
with nonlocal equations of motion?
As in the last section, we may approach this question by looking for
a theory not equivalent to quantum mechanics that obeys relativisitic
causality and nonlocality.$^{13}$
We have considered a model in which action by an
experimenter affects (``jams") nonlocal correlations between
systems measured at spacelike separations from the action.
For example, Shimony$^5$ considers the effect of a laser beam crossing
the path of one of the photons in a singlet pair, after the photon has
already passed.  We find that while nonlocal ``jamming" is not possible
in quantum mechanics, it could be consistent with relativisitic
causality.  If jamming is realized in nature, then perhaps, as suggested
by Grunhaus, it is possible to jam nonlocal quantum correlations.

     We briefly summarize the model.$^{14}$  Two experimenters, call them
Alice and Bob, make measurements on systems that have locally interacted
in the past.  Alice's measurements are spacelike separate from Bob's.
A third experimenter, Jim (the jammer), presses a button on a black box.
This event is spacelike separate from Alice's measurements and from Bob's.
The black box acts at a distance on the correlations between the two sets
of systems.  We find no conflict with relativistic causality if jamming
satisfies two conditions.  The {\it unary} condition requires that
neither Alice, from her results alone, nor Bob, from his, can tell
whether Jim has pressed the button.  Then jamming cannot carry a signal
to either Alice or Bob.  The unary condition implies
indeterminism.  The {\it binary} condition restricts the
range of jamming.  If $A$ and $B$
denote Alice's and Bob's measurements, and $J$ Jim's pressing of the button,
the overlap of the forward light cones
of $A$ and $B$ must lie entirely within the forward light cone of $J$.
\goodbreak
\noindent 4. \hskip 0.5cm SUMMARY
\nobreak

     We have seen that quantum mechanics is not the only theory
combining relativistic causality with nonlocality, nor even, in a sense,
the most nonlocal one.  We found that both ``superquantum" correlations and
a model for nonlocal ``jamming"---a stronger form of
nonlocality than arises in quantum mechanics---can be consistent with
relativistic causality.  The question remains, from what minimal
set of physical principles can we derive quantum mechanics?
\goodbreak
\bigskip
{\bf \quad Acknowledgement.}
\nobreak
The research of D. R. was supported by the State of Israel, Ministry of
Immigrant Absorption, Center for Absorption in Science, and by the Ticho Fund.
\goodbreak
\bigskip
\noindent REFERENCES
\medskip
\nobreak

\noindent
1.  A. Einstein, B. Podolsky, and N. Rosen, {\it Phys. Rev.} {\bf 47}
 (1935) 777.

\noindent
2.  J. S. Bell, {\it Physics} {\bf 1} (1964) 195.

\noindent
3.  Y. Aharonov and D. Bohm, {\it Phys. Rev.} {\bf 115} (1959) 485,
reprinted in F. Wilczek (ed.) {\it Fractional Statistics and Anyon
Superconductivity}, Singapore:  World-Scientific, 1990.

\noindent
4.  It is true that the electron interacts {\it locally} with
a vector potential.  However, the vector potential is not a physical
quantity; all physical quantities are gauge invariant.

\noindent
5.  A. Shimony, in {\it Foundations  of Quantum Mechanics in
Light of the New Technology}, S. Kamefuchi {\it et al.}, eds. (Tokyo,
Japan Physical Society, 1983), p. 225.

\noindent
6.  A. Shimony, in {\it Quantum Concepts of Space and
Time}, R. Penrose and C. Isham, eds. (Oxford, Claredon Press, 1986), p. 182.

\noindent
7.  Y. Aharonov, unpublished lecture notes.

\noindent
8.  S. Popescu and D. Rohrlich, {\it Found. Phys.} {\bf 24},
379 (1994).

\noindent
9.  J. F. Clauser, M. A. Horne, A. Shimony, and R. A. Holt, {\it Phys. Rev.
Lett.} {\bf 23} (1969) 880.

\noindent
10.  B. S. Tsirelson (Cirel'son), {\it Lett. Math. Phys.} {\bf 4} (1980) 93;
L. J. Landau, {\it Phys. Lett.} {\bf A120} (1987) 52.

\noindent
11.  For the maximal violation of the CHSH inequality
consistent with relativity see also L. Khalfin and B. Tsirelson,
in {\it Symposium on the Foundations of Modern Physics '85}, P. Lahti
{\it et al.}, eds. (World-Scientific, Singapore, 1985), p. 441; P. Rastall,
{\it Found. Phys.} {\bf 15}, 963 (1985); S. Summers and R. Werner, {\it J.
Math. Phys.} {\bf 28}, 2440 (1987); G. Krenn and K. Svozil, preprint
(1994) quant-ph/9503010.

\noindent
12.  Y. Aharonov, H. Pendleton, and A. Petersen, {\it Int. J. Theo.
Phys.} {\bf 2} (1969) 213; {\bf 3} (1970) 443; Y. Aharonov, in {\it Proc.
Int. Symp. Foundations of Quantum Mechanics}, Tokyo, 1983, p. 10.

\noindent
13. S. Popescu and D. Rohrlich, in preparation.

\noindent
14. J. Grunhaus, S. Popescu and D. Rohrlich, in preparation.
\bye